\documentclass[conference,letterpaper]{IEEEtran}

\usepackage{amsfonts}
\usepackage{amsmath}
\usepackage{latexsym}
\usepackage{epsfig}
\usepackage{moreverb}
\usepackage{graphics, graphicx,wrapfig}
\usepackage{fancybox}
\usepackage{picinpar,varioref,floatflt}
\usepackage{longtable}
\usepackage{textcomp}
\usepackage{float}
\usepackage{url}
\usepackage{forest}
\usepackage{listings, lstautogobble}
\usepackage{xcolor}
\usepackage{colortbl}
\usepackage{xspace}
\usepackage{siunitx}
\usepackage{multirow}
\usepackage{lscape}
\usepackage{booktabs}
\usepackage{multicol}
\usepackage{array}
\usepackage{tabularx} 
\usepackage{algorithm}
\usepackage{algpseudocode}
\usepackage{subcaption}
\usepackage{makecell}
\usepackage{hhline}
\definecolor{Gray}{gray}{0.95}
\newcolumntype{g}{>{\columncolor{Gray}}p}

\DeclareRobustCommand*{\IEEEauthorrefmark}[1]{%
  \raisebox{0pt}[0pt][0pt]{\textsuperscript{\footnotesize #1}}%
}

\begin{document}

\title{Assessing the Sustainability and Trustworthiness \\ of Federated Learning Models}

\author{
    \IEEEauthorblockN{Chao Feng\IEEEauthorrefmark{1}, Alberto Huertas Celdrán\IEEEauthorrefmark{1,2}, Pedro Miguel S\'anchez S\'anchez\IEEEauthorrefmark{2}, Lynn Zumtaugwald\IEEEauthorrefmark{1}, \\ G\'er\^ome Bovet\IEEEauthorrefmark{3}, Burkhard Stiller\IEEEauthorrefmark{1}}
    
    \IEEEauthorblockA{\IEEEauthorrefmark{1}Communication Systems Group, Department of Informatics, University of Zurich UZH, CH--8050 Zürich, Switzerland \\{[cfeng, huertas, stiller]}@ifi.uzh.ch, lynn.zumtaugwald@uzh.ch}
    \IEEEauthorblockA{\IEEEauthorrefmark{2}Department of Information and Communications Engineering, University of Murcia, 30100 Murcia, Spain \\ pedromiguel.sanchez@um.es}
    \IEEEauthorblockA{\IEEEauthorrefmark{3}Cyber-Defence Campus, armasuisse Science \& Technology, CH--3602 Thun, Switzerland \\ gerome.bovet@armasuisse.ch}
}

\maketitle

\begin{abstract}
Artificial intelligence (AI) increasingly influences critical decision-making across sectors. Federated Learning (FL), as a privacy-preserving collaborative AI paradigm, not only enhances data protection but also holds significant promise for intelligent network management, including distributed monitoring, adaptive control, and edge intelligence. Although the trustworthiness of FL systems has received growing attention, the sustainability dimension remains insufficiently explored, despite its importance for scalable real-world deployment. To address this gap, this work introduces sustainability as a distinct pillar within a comprehensive trustworthy FL taxonomy, consistent with AI-HLEG guidelines. This pillar includes three key aspects: hardware efficiency, federation complexity, and the carbon intensity of energy sources. Experiments using the FederatedScope framework under diverse scenarios, including varying participants, system complexity, hardware, and energy configurations, validate the practicality of the approach. Results show that incorporating sustainability into FL evaluation supports environmentally responsible deployment, enabling more efficient, adaptive, and trustworthy network services and management AI models.

\end{abstract}


\section{Introduction}

As Artificial Intelligence (AI) systems become increasingly embedded in critical infrastructures, including communication networks and cloud-edge services, their role in intelligent network management, such as dynamic resource allocation, traffic optimization, and fault detection, has grown significantly \cite{he2020deployment}. Deep Learning (DL), as the backbone of many AI systems, entails intensive computational demands during training and inference, resulting in substantial carbon emissions. In large-scale networked environments, such as edge-cloud infrastructures and distributed systems, this resource consumption increases with the number of participating agents, posing a growing sustainability concern. Moreover, DL’s dependence on massive datasets often leads to inefficient data handling, adding further energy overhead. These challenges are particularly prominent in network management tasks, where AI is employed for dynamic resource allocation, traffic control, and fault diagnosis across heterogeneous infrastructures.

Sustainability in AI extends beyond energy and computation. Ethical concerns, such as bias, privacy violations, and discrimination, are especially critical in decentralized and privacy-sensitive domains like personalized network services. Thus, sustainable AI must be pursued in conjunction with broader pillars of trustworthiness, including robustness, transparency, fairness, and accountability.

To this end, ensuring AI trustworthiness has become a global priority. Regulatory initiatives, such as those led by the European Commission’s High-Level Expert Group on AI (AI-HLEG) \cite{AI_HLEG}, have established comprehensive guidelines. The AI-HLEG outlines seven core requirements for trustworthy AI: human agency and oversight, technical robustness and safety, privacy and data governance, transparency, fairness, environmental well-being, and accountability \cite{kaur2022trustworthy}.

As highlighted by the AI-HLEG, data privacy is a challenging and active research topic within trustworthy AI. In 2016, Google introduced Federated Learning (FL) \cite{mcmahan2017communication}, an innovative paradigm that enables multiple clients to collaboratively train models without necessitating the exchange of private data. Today, FL confronts multifaceted challenges, spanning scalability, single point of failure, architectural design, or privacy and security concerns, among others. However, while FL inherently incorporates privacy-preserving features, trustworthy AI remains a pivotal dimension within FL systems \cite{sanchez2023federatedtrust}. 

\begin{table*}[h!]
\centering
\caption{Existing Trustworthy FL Taxonomies and Their Coverage of Pillars and AI-HLEG Requirements}
\label{tab:ftaxonomies}
\resizebox{\textwidth}{!}{%
\setlength\arrayrulewidth{1pt}
\begin{tabular}{p{1.7cm}p{1.5cm}p{2cm}p{2cm}p{2cm}p{2.5cm}p{2cm}g{2cm}}
\hline
\rowcolor{white}
 \textit{Authors} &
  \multicolumn{6}{c}{\textit{Pillars/AI-HLEG Requirements}} &
\\ \hhline{~-------} 
  \textit{(Year)}&
  Privacy &
  Fairness &
  Robustness &
  Accountability &
  Explainability &
  Federation &
  Sustainability \\ \hhline{~-------} 
  &
  3. Privacy and data governance &
  5. Diversity, non-discrimination, and fairness &
  2. Technical robustness and safety &
  7. Accountability and auditability / 1. Human agency and oversight &
  4. Transparency including explainability &
  2. Technical robustness and safety / 5. Diversity, non-discrimination and fairness &
  6. Environmental well-being \\ \hline
Shi et al. \cite{shi2021survey} (2021)&
  No &
  Yes &
  No &
  No &
  No &
  Partially &
  No \\ \hline
  Liu et al. \cite{liu2022trustworthy} (2022)&
  Yes &
  No &
  Yes &
  No &
  No &
  Partially &
  No \\ \hline
Tariq et al. \cite{tariq2023trustworthy} (2023)&
  Yes &
  Yes &
  Yes &
  No &
  Yes &
  No &
  No \\ \hline
    Zhang et al. \cite{zhang2023survey} (2023)&
    Yes &
    No &
    Yes &
    No &
    No &
    No &
    No
    \\
  \hline
Sanchez et al.\cite{sanchez2023federatedtrust} (2023)&
  Yes &
  Yes &
  Yes &
  Yes &
  Yes &
  Yes &
  No \\ \hline
  \textbf{This work}&
  Yes &
  Yes &
  Yes &
  Yes &
  Yes &
  Yes &
  Yes \\ \hline
\end{tabular}%
}
\end{table*}

In this context, prior work \cite{celdran2023framework,tariq2023trustworthy} defined a baseline by formulating taxonomies for trustworthy ML, DL, and FL. Other work, such as \cite{sanchez2023federatedtrust}, implemented algorithms and frameworks for assessing the trustworthiness of FL systems. However, environmental well-being is missing in full there. Specifically, key sustainability factors such as CO$_2$eq emissions, hardware efficiency, federation complexity, and energy grid carbon intensity, highlighted by AI-HLEG, remain unaddressed, despite their importance for optimizing FL configurations and promoting awareness. Moreover, unlike centralized paradigms, FL operates in a decentralized manner, involving heterogeneous and geographically distributed clients. This results in substantial variability in computational capabilities and local carbon intensities across nodes, making sustainability assessment more complex. FL's dynamic and distributed nature increases the challenge of designing effective, context-aware environmental impact evaluation metrics.

This work makes three key contributions toward sustainable and trustworthy Federated Learning (FL). First, it proposes a unified taxonomy for Trustworthy FL with seven pillars: privacy, robustness, fairness, accountability, federation, explainability, and a newly introduced sustainability pillar, which encompasses carbon intensity, hardware efficiency, and federation complexity, along with ten corresponding metrics. Second, it presents an evaluation algorithm (available in \cite{github}) that incorporates sustainability metrics into the trustworthiness assessment. Third, the algorithm is integrated into the FederatedScope framework and validated across diverse settings, demonstrating its effectiveness in evaluating FL trustworthiness. By explicitly addressing sustainability, the proposed approach supports more efficient and responsible FL-based network management approaches, including intelligent orchestration, edge coordination, and energy-aware optimization.

\section{Related Work}
\label{sec:related}

This section reviews recent and relevant work documented in the literature regarding trustworthy FL evaluation and carbon emission estimation for AI/FL-based computing.

\subsection{Trustworthy FL Evaluation}

\tablename~\ref{tab:ftaxonomies} summarizes the existing trustworthy FL taxonomies and their coverage of trustworthy FL pillars defined by the AI-HLEG. The taxonomy from Shi et al. \cite{shi2021survey} reviewed the issue of fairness in FL and its evaluation mechanisms. This study only covers the pillar of fairness and partially the federation one, since it discusses fair client selection. Liu et al. \cite{liu2022trustworthy} provided a taxonomy covering the pillars of privacy, robustness, and partially the pillar of federation. Tariq et al. \cite{tariq2023trustworthy} proposed an architecture for FL trustworthiness. Its taxonomy covers privacy, fairness, explainability, and robustness pillars and includes requirements two, three, and five defined by the AI-HLEG. Zhang et al. \cite{zhang2023survey} also surveyed trustworthy FL, but focusing on the legal aspects of security, privacy, and robustness pillars. The taxonomy that covers the most pillars and requirements defined by the AI-HLEG is the trustworthy FL taxonomy from S\'anchez et al. \cite{sanchez2023federatedtrust}. The taxonomy contains the pillars i) privacy, ii) robustness, iii) fairness, iv) explainability, v) accountability, and vi) federation. For each pillar, notions and metrics are defined. In total, 36 metrics are defined that can be used to evaluate the trustworthiness score of FL. 

After reviewing the literature, the most important limitation becomes apparent when comparing the taxonomy to the requirements defined by the AI-HLEG and the existing taxonomies. The environmental impact of an FL system is not considered in the taxonomy, but environmental well-being has clearly been defined as one of the seven requirements for trustworthy AI by governing bodies \cite{AI_HLEG}. Since \cite{sanchez2023federatedtrust} is the most advanced taxonomy that covers six of the seven requirements defined by the AI-HLEG, it is employed as the basis for extension, considering the environmental impact of the system.

\subsection{Estimating Emissions of AI/FL}
Most works focus on estimating the carbon emissions of specific models. Lucconi et al. \cite{luccioni2023counting} provided a survey on aspects that influence the CO$_2$eq of ML. Luccioni et al. Patterson et al. \cite{patterson2021carbon} estimated the energy consumption and computed the carbon emissions of the language models T5, Meena, GShard, Switch Transformer, and GPT-3 and highlighted opportunities to improve energy efficiency and CO$_2$eq emission, such as sparsely activated DNNs and using energy grids with low carbon intensity. In the field of FL, Qui et al. \cite{qiu2023first} provided a first look into the carbon footprint of FL models by incorporating parameters specific to FL and comparing the emissions produced by FL models with those produced by centralized ML models. Similarly to estimating the carbon emissions of AI/FL models, tools have been developed to track carbon emissions. CodeCarbon \cite{codecarbon} and the Experimental Emissions Tracker \cite{henderson2020systematic} can be used to track emissions during the training process, while the ML CO$_2$eq Calculator \cite{lacoste2019quantifying} can be used to calculate the emissions after training. 

\begin{figure*}[!ht]
\centerline{\includegraphics[width=0.7\textwidth]
{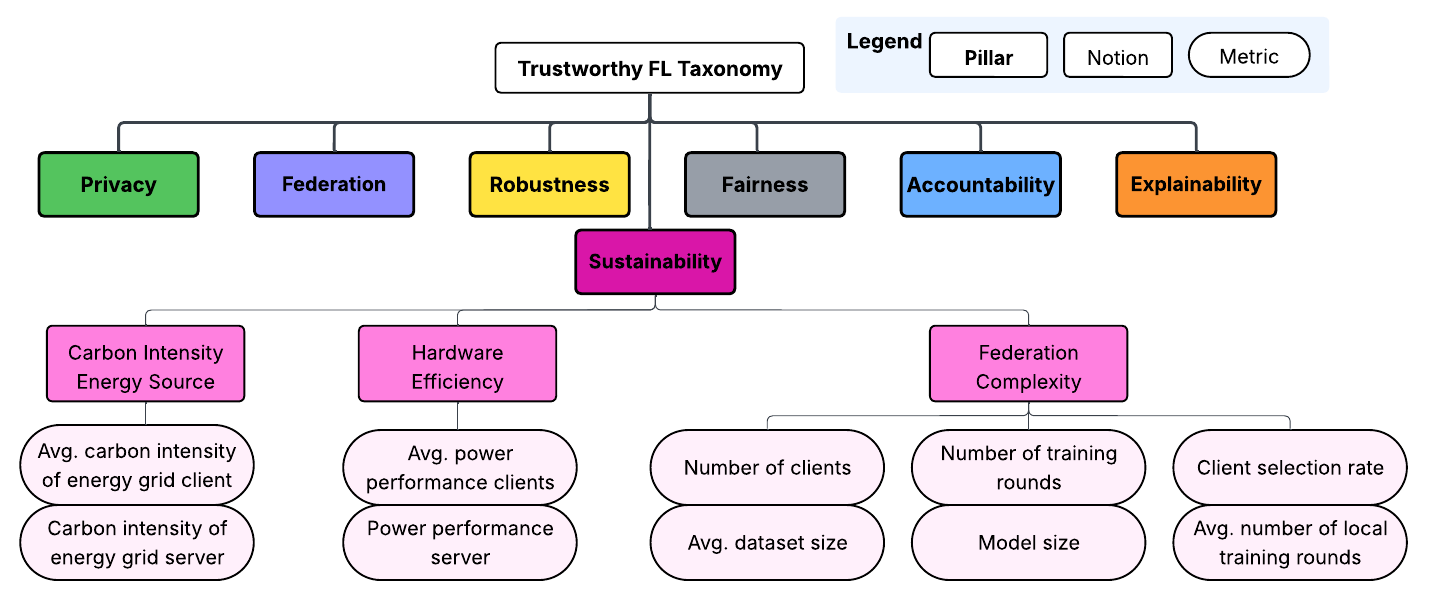}}
\caption{Trustworthy FL Taxonomy}
\label{FL_taxonomy_new}
\end{figure*}

Despite the effort and work done in this research field, none have incorporated the emissions produced by FL models into trustworthy FL, despite environmental well-being clearly being defined as one of the seven key requirements for trustworthy AI/FL by the AI-HLEG \cite{AI_HLEG}.

\section{The Sustainability Pillar of Trustworthy FL}
\label{sec:sustainability}
The trustworthy FL taxonomy is structured hierarchically into three levels. At the top level, \textbf{pillars} represent the fundamental aspects to be considered, such as privacy, robustness, and sustainability. Each pillar is decomposed into several \textbf{notions}, which capture the specific dimensions that need to be addressed to achieve the corresponding pillar. Finally, each notion is associated with one or more \textbf{metrics} that allow its quantitative assessment.

This section describes the notions and metrics that comprise trustworthy FL's sustainability pillar. This pillar includes the energy grid's carbon intensity, the underlying hardware's efficiency, and the federation's complexity. The overview of the Trustworthy FL framework is presented in \figurename~\ref{FL_taxonomy_new}. 

\subsection{Carbon Intensity}
Carbon intensity depends on the energy source used for electricity generation. The energy mix used to train FL models significantly impacts total emissions. For example, training with 500 kWh from coal results in 410 kg CO$_2$eq, versus just 5.5 kg with nuclear power. National grids vary widely, ranging from 20g in Lesotho to 795g CO$_2$eq/kWh in Botswana \cite{petroleum2022statistical}. Therefore, this notion seeks to measure the carbon impact of FL according to the following two metrics.

\begin{itemize}
    \item \textbf{Client/Server Carbon Intensity}. These two metrics measure the carbon intensity of the energy grid utilized in the FL process from the perspectives of both the clients and the server. The value of these two metrics ranges from 20g of CO$_2$eq to 795 of CO$_2$eq by looking at the countries' energy grids, according to the IPCC report \cite{schlomer2014annex}. Theoretically, with the energy sources available today, the lowest possible energy grid would have 11g of CO$_2$eq per kWh only using wind energy and the highest possible 820g of CO$_2$eq only using coal energy. The energy grids used by clients can be determined by the location of the federation clients (retrieved from the IP address). The carbon intensity of the energy grid utilized by clients is determined by calculating the average of all the carbon intensities. For the carbon intensity of the energy grid used by the server, the energy grid of the country the server operates in is taken. Equation \ref{eq:carbon} illustrates the calculation process of this metric.
    \begin{equation}
    T_{Intensity}=S_{Intensity} + \frac{1}{n} \sum_{i=1}^{n} {C_n}_{Intensity}
    \label{eq:carbon}    
    \end{equation}
    Where $T_{Intensity}$ represents the total grid carbon energy intensity, $S_{Intensity}$ represents the server grid carbon intensity, and ${C_n}_{Intensity}$ represents the grid carbon intensity of each client $n$.
\end{itemize}

\subsection{Hardware Efficiency}
The second notion that significantly impacts the energy consumption and, thus, the emissions of an FL system is the efficiency of the underlying hardware. Efficient hardware consumes less power to perform computational tasks. Lower power consumption translates to reduced energy requirements, leading to lower CO$_2$eq emissions. On the contrary, inefficient hardware generates more heat, necessitating additional cooling mechanisms, such as air conditioning or fans, that contribute to more CO$_2$eq emissions \cite{lacoste2019quantifying}. In FL systems, both the process of training local models and the aggregation of these models globally require heavy computational resources. Thus, the efficiency of the underlying hardware plays a significant role in the emissions produced by the FL system. 

The performance of CPUs and GPUs can be described by different metrics, such as clock speed, Floating-Point Operations Per Second, or Instructions Per Second (IPS) \cite{martonosi2001modeling}. It is important to note that none of these metrics provides a complete picture of the performance of the processing units, and different metrics are more relevant in certain use cases. Further, manufacturers of CPUs and GPUs often do not fully disclose the metrics of their products, which makes comparing them difficult. To solve this issue, lots of benchmarking software to evaluate the processor's performance across a range of tasks has been proposed. In terms of heat production of a processor, Thermal Design Power (TDP) is used as a specification in the industry \cite{PassMark}. It indicates the maximum amount of heat a computer component, such as a CPU or GPU, is expected to generate under normal operating conditions. TDP is typically expressed in watts and represents the maximum power consumption and heat dissipation expected under typical workloads. The smaller the number for TDP, the lower the power consumption of the processor. Therefore, the Hardware Efficiency notion proposes the following metrics.

\begin{itemize}
    \item \textbf{Client/Server Hardware Efficiency}. To evaluate the efficiency of the underlying hardware in terms of computing power per unit of power consumed, it makes sense to divide the benchmark performance through the TDP, defining the power performance of the processor. A processor with a high power performance score is able to do a lot of computation with low energy consumption, and it is thus more efficient in terms of resource consumption \cite{PassMark}. It is measured in performance per Watt using Equation \ref{eq:efficiency} and \ref{eq:hardware}.
    \begin{equation}
    H_{E}=\frac{H_{BP}}{H_{TDP}}
    \label{eq:efficiency}    
    \end{equation}
    \begin{equation}
    Total_{E}=S_{E} + \frac{1}{n} \sum_{i=1}^{n} {C_n}_{E}
    \label{eq:hardware}    
    \end{equation}
    Where $H_{E}$ is the hardware efficiency score, $H_{BP}$ is the hardware benchmark performance, $H_{TDP}$ is the hardware TDP, $S_{E}$ is the server hardware efficiency, and ${C_n}_{E}$ is the hardware efficiency of each client $n$.
\end{itemize}

\subsection{Federation Complexity}

The complexity and size of the federation impact the consumed energy and, thus, the emissions produced. Generally, the more complex the model, the higher the number of participants and the higher the energy consumption \cite{qiu2023first}. Therefore, the federation complexity notion considers the following metrics.

\begin{itemize}

    \item \textbf{Number of Training Rounds}. More rounds increase client/server computation and communication energy.
    
    \item \textbf{Dataset Size}. Larger datasets demand more memory, compute, and energy \cite{lacoste2019quantifying}.
    
    \item \textbf{Model Size}. Large models typically require more computational resources and time to process each iteration, which results in higher energy consumption \cite{lacoste2019quantifying} at the client's side. Also, aggregating large models on the server side typically uses more energy than aggregating small models due to the number of weights. Furthermore, large models also introduce a communication overhead. 

    \item \textbf{Number of Clients}. The more clients participate in the federation, the more energy is used \cite{qiu2023first} for i) training, ii) aggregation, and iii) communication, and thus, the more CO$_2$eq are emitted. 

    \item \textbf{Client Selection Rate}. The larger this percentage, the larger the communication overhead from the uplink communication, and the larger the CO$_2$eq emissions.

    \item \textbf{Number of Local Training Rounds}. The higher the number of local training rounds, the higher the computational overhead on the client's side and the higher the energy consumption \cite{lacoste2019quantifying, qiu2023first}. 
    
\end{itemize}

\subsection{Additional Pillars of Trustworthy FL}

The six pillars defined by S\'anchez et al. \cite{sanchez2023federatedtrust} together with the new one cover the seven requirements for trustworthy AI defined by the AI-HLEG \cite{AI_HLEG} and constitute a comprehensive taxonomy.

\subsubsection{Privacy}
While FL offers inherent privacy benefits, it assumes honest behavior from participants. To mitigate risks from curious or malicious actors, this pillar considers four aspects: adoption of privacy-preserving methods, metrics for quantifying information leakage, and the likelihood of inferring private data from client updates.

\subsubsection{Robustness}
Ensuring robustness is essential to protect AI systems from adversarial threats and failures. This includes: resilience to adversarial attacks, robustness of hardware/software used by participants, and reliability of FL algorithm performance and customization.

\subsubsection{Fairness}
Fairness in FL is challenged by heterogeneous client data. This pillar includes client selection fairness, group-level fairness (non-discrimination), and individual-level fairness through performance alignment and label distribution balance across clients.

\subsubsection{Explainability}
Explainability ensures transparency in AI systems. It includes two aspects: intrinsic model interpretability and post-hoc explainability methods, which are particularly important in FL due to privacy constraints that limit access to raw data.

\subsubsection{Accountability}
Accountability is addressed through FactSheet Completeness, documenting the ML pipeline, and Monitoring, which ensures participants adhere to procedural and architectural standards throughout the FL model lifecycle.

\subsubsection{Federation}
This pillar tackles challenges in FL coordination, such as communication overhead, resource limitations, and client heterogeneity. Key aspects include client/model management and the design of optimization algorithms to ensure stable and efficient training.

\begin{table*}[!ht]
\small
\centering
\scriptsize
\caption{Metrics for Sustainability Pillar}
\label{tab:metrics_sus_pillar}
\begin{tabular}{ p{2.4cm}p{5.2cm}p{2.3cm}p{1.5cm}p{4.5cm}}
\toprule
\textit{\textbf{Metric}} &
  \textit{\textbf{Description}} &
  \textit{\textbf{Input}} &
  \textit{\textbf{Output}} &
  \textit{\textbf{Normalized Output}} \\ \hline
   \hline
\multicolumn{5}{c}{\textit{\textbf{Notion: Carbon Intensity of Energy Source}}} \\
 \hline
Avg, carbon intensity of clients &
  Average carbon intensity of energy grid used by clients &
  Location of clients (IP) &
  Float {[}20,795{]} & 
  ${(795-output)}/{(795-20)}$  \\
   \hline
Carbon intensity server &
  Carbon intensity of energy grid used by the server &
  Location of server (IP) &
  Float {[}20,795{]} &
   ${(795-output)}/{(795-20)}$   \\
   \hline
   \hline
\multicolumn{5}{c}{\textit{\textbf{Notion: Hardware Efficiency}}} \\
   \hline
Avg. hardware efficiency of clients &
  Average performance per watt (CPU or GPU Mark/ TDP) of CPUs and GPUs used by clients &
  CPU and GPU models of clients &
  Float {[}20,1447{]} &
     ${(1447-output)}/{(1447-20)}$   \\
     \hline
Hardware efficiency of clients &
  Performance per watt (CPU or GPU Mark/ TDP) of CPUs and GPUs used by the server &
  CPU and GPU models of server &
  Float {[}20,1447{]} &
     ${(1447-output)}/{(1447-20)}$   \\
     \hline   \hline
\multicolumn{5}{c}{\textit{\textbf{Notion: Federation Complexity}}} \\
   \hline
Number of  global training rounds &
  Number of global training rounds in the FL system &
  Config file &
  Integer &
  \makecell[l]{output:$[$10, $10^2$, $10^3$, $10^4$, $10^5$, $10^6$$]$\\norm:$[$1, 0.8, 0.6, 0.4, 0.2, 0$]$} \\
     \hline
Number of  clients &
  Number of clients in the federation &
  Config file &
  Integer &
  \makecell[l]{output:$[$10, $10^2$, $10^3$, $10^4$, $10^5$, $10^6$$]$\\norm:$[$1, 0.8, 0.6, 0.4, 0.2, 0$]$} \\
     \hline
Client selection rate &
  Percentage of clients selected in each training round to share their models &
  Config file &
  Float {[}0,1{]} &
  {[}0,1{]} \\
     \hline
Average number of local training rounds &
  Average number of local training rounds performed by clients within one global training round &
  Config file &
  Integer &
  \makecell[l]{output:$[$10, $10^2$, $10^3$, $10^4$, $10^5$, $10^6$$]$\\norm:$[$1, 0.8, 0.6, 0.4, 0.2, 0$]$} \\
     \hline
Average dataset size &
  Average number of samples used by clients in one training round &
  Client Statistics &
  Integer &
   \makecell[l]{output:$[$$10^5$, $10^6$, $10^7$, $10^8$, $10^9$, $10^{10}$$]$\\norm:$[$1, 0.8, 0.6, 0.4, 0.2, 0$]$}\\
     \hline
Model size &
  Number of features/depth of decision tree/number of parameters in NN &
  Model & Integer
   &
   \makecell[l]{output:$[$$10^5$, $10^6$, $10^7$, $10^8$, $10^9$, $10^{10}$$]$\\norm:$[$1, 0.8, 0.6, 0.4, 0.2, 0$]$}\\ \bottomrule
\end{tabular}
\end{table*}

\section{Sustainable and Trustworthy FL Algorithm}
\label{sec:design}

This section provides the details of the algorithm in charge of assessing the sustainability and trustworthiness of FL models. The main contribution of this algorithm, compared to the literature, is the design and implementation of three notions and ten metrics dealing with the sustainability pillar and their integration with six other existing pillars (privacy, robustness, fairness, accountability, federation, and explainability). The following assumptions (A), functional requirement (FR), non-functional requirements (NF), and privacy constraint (PC) were considered during the algorithm design phase. 

\begin{itemize}
   
    \item A\_1: The central server is honest. It is maintained by a trusted owner, and it does not interfere with the FL protocol maliciously.

    \item A\_2: Clients of the federation are honest but curious. They trustfully report their metrics and statistics without maliciously interfering with the FL protocol.
    
    \item FR\_1: The three notions and ten metrics of the Sustainability pillar must be represented in the algorithm. In addition, each of the remaining six trustworthy FL pillars must be considered, meaning that at least one metric from each pillar must be considered in the final score.
    
    \item FR\_2: The final trustworthiness score must be a combination of the trustworthiness scores from all notions and pillars. 
    
    \item NF\_1: The algorithm should add minimal computation overhead and complexity to the server, participants, and FL model.
    
    \item NF\_2: The algorithm should be modular and configurable.

    \item PC\_1: The algorithm must not store sensitive data from the FL model.
    
    \item PC\_2: The algorithm must not leak or share sensitive data from clients, the server, and the FL model with third parties.
    
    \item PC\_3: The metrics calculations can occur at the client’s local devices, the central server, or collaboratively between both.
    
\end{itemize}

\subsection{Sustainability Pillar: Notions and Metrics}

\tablename~\ref{tab:metrics_sus_pillar} shows the notions and metrics explained in Section~\ref{sec:sustainability} and considered in the algorithm for the sustainability pillar. Descriptions, inputs, outputs, and normalization details are provided for each metric. For metric computation, the CodeCarbon package \cite{codecarbon} is leveraged to obtain the emissions related to the hardware employed by the server/clients and the emissions related to the location of the nodes in the FL setup. This package has been selected by the most representative solutions in the literature, as described in Section~\ref{sec:related}. Besides, for the calculation of \textit{Hardware Efficiency metrics}, the most popular benchmarking software for processors is PassMark \cite{PassMark}. It computes a performance score by running standardized tests that simulate real-world workloads, such as executing complex mathematical calculations. PassMark has provided a database with Power Performance measurement for over 3000 CPUs and 2000 GPUs published on Kaggle, which can be used to evaluate the client and server processor efficiency in the algorithmic prototype design.

In addition to the previous ten metrics, the proposed algorithm also implements the 41 metrics belonging to the remaining six pillars proposed in  \cite{sanchez2023federatedtrust}.

\subsection{Algorithm Design}
 The proposed algorithm considers the following inputs.

\begin{itemize}

    \item \textit{Emissions}. It contains the IP of clients and server, CPU and GPU models, and config files of the federation needed to compute the ten sustainability metrics (see \tablename~\ref{tab:metrics_sus_pillar}).

    \item \textit{FL Model}. It contains information about the model configuration and model personalization.
    
    \item \textit{FL Framework Configuration}. It contains information about the number of clients, the client selection mechanisms, the aggregation algorithm, and the model hyperparameters. 
    
    \item \textit{FactSheet}. It contains essential details for the accountability of the training process, federation, and the individuals involved \cite{arnold2019factsheets}.
    
    \item \textit{Statistics}. It contains information about the client class balance, client test performance loss, client test accuracy, client clever score, client feature importance, client participation rate, client class imbalance, client average training time, model size, and average upload/download bytes.
    
\end{itemize}

These input sources serve as the foundation for deriving the sustainability metrics outlined in \tablename~\ref{tab:metrics_sus_pillar} and the metrics belonging to the remaining six pillars proposed in \cite{sanchez2023federatedtrust}. The resulting metric values are then normalized to ensure a consistent range. It is essential to note that each metric can encompass distinct input sources and may be computed at different stages of the federated learning (FL) model creation process, namely pre-training, during-training, or post-training, by various participants within the federation, be it clients or servers. Once the normalized metric outputs are determined, they are assigned weights and combined to produce a score for each notion. Each pillar incorporates one or more notions, assessed based on predefined yet adjustable weights for each metric. Consequently, the same procedure is reiterated to derive pillar scores through the weighting and aggregation of notion scores. Ultimately, the overall trust score of the FL model is determined as a custom amalgamation of the pillar scores. 

\subsection{Algorithm Deployment}

Once designed, the algorithm was implemented and deployed in a well-known FL framework called FederatedScope \cite{xie2022federatedscope}. After the deployment, the following steps show how the sustainability and trustworthy FL scores are calculated.

\begin{enumerate}
    \item \textit{Setup}: FederatedScope initializes clients, server, and the trustworthiness algorithm using the given configuration, populating the FactSheet with pre-training metrics.
    
    \item \textit{Model Broadcast}: The server sends the global model to selected clients.
    
    \item \textit{Local Training}: Clients train locally and track sustainability metrics using CodeCarbon.
    
    \item \textit{Report Emissions}: Clients report hardware and energy grid info, stored in the emissions file.
    
    \item \textit{Model Sharing}: Clients send updated model parameters to the server.
    
    \item \textit{Aggregation}: The server securely aggregates client updates.
    
    \item \textit{Evaluation}: Clients evaluate the model and trigger metric computation.
    
    \item \textit{Iteration}: Steps 2–7 repeat for all training rounds.
    
    \item \textit{Finalize Results}: Final evaluation results are added to the FactSheet.
    
    \item \textit{Trust Score}: The algorithm computes the final trust score and generates the report.
\end{enumerate}

The execution of the FederatedScope training process, together with the evaluation of the FL sustainability and trustworthiness, is depicted in Algorithm \ref{alg:ft_reduced}.

\begin{algorithm}[!ht]
\small
\caption{Training in FederatedScope (Reduced)}\label{alg:ft_reduced}
\hspace*{\algorithmicindent} \textbf{Input:} $N$ clients, sample size $m$, server $S$, iterations $T$, initial model $\overline{w}(0)$, config $C$, FederatedTrust manager $ft$ \\
\hspace*{\algorithmicindent} \textbf{Output:} Evaluation results, trust report, carbon estimates
\begin{algorithmic}[1]
    \State $S$ shares client IDs and config $C$ with $ft$; $ft$ initializes FactSheet and metrics
    \State $S$ requests class distribution; $ft$ builds emission file and class map
    \For{each client $i \in [N]$} 
        \State Client $i$ reports class stats to $ft$
    \EndFor
    \For{$t = 0$ to $T$}
        \State $S$ samples $m$ clients $D(t)$ and notifies $ft$
        \State $S$ broadcasts model $\overline{w}(t)$
        \For{each client $i \in D(t)$}
            \State Track emissions, train locally, update $ft$, send $w_i(t+1)$ to $S$
        \EndFor
        \State $S$ tracks emissions, aggregates updates to $\overline{w}(t+1)$, updates $ft$
    \EndFor
    \State $S$ sends final model $\overline{w}'$ to all clients
    \For{each client $i \in [N]$}
        \State Evaluate $\overline{w}'$ locally, send results to $S$
    \EndFor
    \State $S$ aggregates results, sends to $ft$; $ft$ finalizes report and trust score
\end{algorithmic}
\end{algorithm}

\section{Evaluation and Results}
\label{sec:evaluation}

This section evaluates the proposed algorithm through a pool of experiments. Firstly, it includes a quantitative analysis of its functionality. Then, it analyzes how the proposed system can effectively help users to better understand the sustainability of the FL systems and support decision-making processes.

\subsection{Functionality Evaluation}
Four use cases (UC) are conducted to examine the functionality of the sustainability pillar. They consider several levels of federation complexity, diverse degrees of carbon intensity in the energy grid utilized by both clients and the server, and different hardware efficiencies of the CPUs employed by the clients and the server. The setups for these four cases are depicted in \tablename~\ref{tab:setup_functionality}. In the following experiments, each metric carries equal weight when calculating the notion score. In addition, when determining the sustainability pillar score, the carbon intensity of the energy source metric is assigned a weight of 0.5, while the hardware efficiency and federation complexity metrics are each assigned a weight of 0.25.

\begin{table}[ht!]
\centering
\caption{Setups for Functionality Evaluation Experiment}
\label{tab:setup_functionality}
\resizebox{\columnwidth}{!}{%
\begin{tabular}{lllll}
\toprule
& \textit{UC A} & \textit{UC B} & \textit{UC C} & \textit{UC D} \\ \midrule
   \hline

Clients Loc. &
  Albania &
  \begin{tabular}[c]{@{}l@{}}50\% in Kosovo\\ 50\% in Gambia\end{tabular} &
  Switzerland &
  South Africa \\ \hline
Server Loc.                         & Albania      & South Africa & Switzerland    & South Africa  \\ \hline
Clients Hardware &
  i7-1250U &
  AMD FX-9590 &
  \begin{tabular}[c]{@{}l@{}}40\%  E5-4620\\ 35\%  E5-4627\\ 25\%  E5-2650\end{tabular} &
  i5-1335U \\ \hline
Server Hardware &
   i7-1250U &
   W2104 &
   E5-4620 &
   i7-1250U \\ \hline 
Rounds               & 10           & 1000     & 1000       & 10            \\ \hline
No. of Clients                       & 5            & 1000     & 1000       & 8             \\ \hline
Selection Rate                   & 0.2          & 1            & 0.8            & 0.3           \\ \hline
Local Rounds & 1            & 90           & 90             & 1             \\ \hline
Dataset Size                    & 100          & 1.10E+06     & 1.10E+06       & 100           \\ \hline
Model size                              & 98,000       & 1.00E+13     & 1.00E+13       & 99,300     \\ \bottomrule  
\end{tabular}%
}
\end{table}

\subsubsection{Low Carbon Intensity and High Hardware Efficiency} UC A represents the optimal situation with minimal CO$_2$eq emissions. In this scenario, the server and all five clients utilize the Intel Core i7-1250U CPU, which boasts exceptional efficiency with a power performance of 1447, the greatest recorded by PassMark thus far. Moreover, the federation complexity remains low, characterized by a limited number of clients, global and local training rounds, as well as a small client selection rate, dataset size, and model size. Furthermore, both the clients and server are situated in Albania, which possesses one of the least carbon-intensive energy grids. Therefore, as depicted in \tablename~\ref{tab:results_functionality}, UC A obtains a carbon intensity of energy source notion score of 1, a hardware efficiency notion score of 1, and a federation complexity notion score of 0.98, resulting in the highest result with an overall sustainability score.

\begin{table}[ht]
\centering
\caption{Sustainability Score for Functionality Evaluation}
\label{tab:results_functionality}
\resizebox{\columnwidth}{!}{%
\begin{tabular}{lllll}
\toprule
\textit{Metric}                                                                              & \textit{UC A} & \textit{UC B} & \textit{UC C} & \textit{UC D} \\ \midrule
   \hline
\textbf{Sustainability Pillar}                                                      & 1.00       & 0.09       & 0.55       & 0.53       \\  \hline
\textbf{\begin{tabular}[c]{@{}l@{}}- Carbon Intensity of  Energy \\  Source Notion (weight 0.5)\end{tabular}} & 1.00 & 0.09 & 1.00 & 0.11 \\ \hline
\begin{tabular}[c]{@{}l@{}}- - Avg. Carbon Intensity of \\ Energy Grid Clients\end{tabular}    & 1.00 & 0.08 & 1.00 & 0.11 \\ \hline
\begin{tabular}[c]{@{}l@{}}- - Carbon Intensity of \\ Energy Grid Server\end{tabular}          & 1.00 & 0.11 & 1.00 & 0.11 \\ \hline
\textbf{\begin{tabular}[c]{@{}l@{}}- Hardware Efficiency \\ Notion (weight 0.25)\end{tabular}}    & 1.00       & 0.01       & 0.04       & 0.94       \\ \hline
\begin{tabular}[c]{@{}l@{}}- - Avg. Hardware \\ Efficiency Clients\end{tabular}     & 1.00       & 0.01       & 0.05       & 0.87       \\ \hline
\begin{tabular}[c]{@{}l@{}}- - Hardware Efficiency \\ Server\end{tabular}           & 1.00       & 0.02       & 0.04       & 1.00       \\ \hline
\textbf{\begin{tabular}[c]{@{}l@{}}- Federation Complexity \\ Notion (weight 0.25)\end{tabular}}  & 0.98       & 0.13       & 0.17       & 0.96       \\ \hline
\begin{tabular}[c]{@{}l@{}}- - Number of  Training \\ Rounds\end{tabular}           & 1.00       & 0.17       & 0.17       & 1.00       \\ \hline
- - Number of Clients                                                                & 1.00       & 0.17       & 0.17       & 1.00       \\ \hline
- - Client Selection Rate                                                           & 0.89       & 0.00       & 0.22       & 0.77       \\ \hline
\begin{tabular}[c]{@{}l@{}}- - Avg. Number of Local \\ Training Rounds\end{tabular} & 1.00       & 0.17       & 0.10       & 1.00       \\ \hline
- - Average Dataset Size                                                            & 1.00       & 0.20       & 0.20       & 1.00       \\ \hline
- - Model Size                                                                      & 1.00       & 0.14       & 0.14       & 1.00  \\ \bottomrule    
\end{tabular}%
}
\end{table}

\subsubsection{High Carbon Intensity and Low Hardware Efficiency}
UC B illustrates a worst-case scenario with inefficient hardware, a highly complex federation, and carbon-intensive electricity grids, resulting in substantial CO$_2$eq emissions. The server uses an Intel Xeon W-2104 CPU (power performance 51.67), and all 1000 clients use AMD FX-9590 CPUs (power performance 30.76), leading to a hardware efficiency score of 0.01. The federation involves 1000 global and 90 local rounds, with a complexity score of 0.13. The server is located in South Africa (709g CO$_2$/kWh), with half of the clients in Kosovo (769g) and the other half in Gambia (700g), averaging 734.5g and resulting in a carbon intensity score of 0.09. Combining these three notions with the weighted average, the overall score for the sustainability pillar is 0.09 for UC B, which represents a worst-case scenario in terms of sustainability with inefficient hardware and carbon-intensive electricity grids in combination with a complex federation.

\subsubsection{Low Carbon Intensity and Low Hardware Efficiency} UC C represents a scenario where the hardware used is inefficient, and the federation is complex, leading to high energy consumption. However, the carbon intensity of the electricity grid is low, resulting in medium CO$_2$eq emissions. In this case, the server utilizes an Intel Core i7-6800K CPU with a power performance of 76.29. Among the clients, 40\% use an Intel Xeon E5-4620 CPU with a power performance of 100.24, 35\% use an Intel Xeon E5-4627 with a power performance of 71.69, and 25\% use an Intel Xeon E5-2650 with a power performance of 105.21. Overall, the hardware is considered inefficient, achieving a hardware efficiency notion score of 0.01. The federation complexity is high, with a large number of clients, global training rounds, local training rounds, and parameters in the DNN model, resulting in a federation complexity notion score of 0.17. However, both the server and clients are located in Switzerland, where the energy grid has a low carbon intensity of 32g CO$_2$eq per kWh, achieving a carbon intensity of energy source notion score of 1. By combining these three notions, the overall score for the sustainability pillar is 0.55 for UC C.

\subsubsection{High Carbon Intensity and High Hardware Efficiency} UC D utilizes highly efficient computational hardware but has a high carbon intensity in its grid, leading to a moderate level of CO$_2$eq emissions, in contrast to UC C. In UC D, the server utilizes the Intel Core i7-1250U CPU, with a power performance of 1447, while all eight clients use the Intel Core i5-1335U, with a power performance of 1268. Additionally, the federation complexity is low, with a small number of clients, global training rounds, local training rounds, and a small client selection rate, dataset size, and model size. Consequently, the hardware efficiency notion score and federation complexity notion score are 0.94 and 0.96, respectively. However, both the clients and server are situated in South Africa, where the carbon intensity of the energy source is 709g CO$_2$eq per kWh, resulting in a carbon intensity of energy source notion score of 0.11. Therefore, the final score is 0.53, similar to UC C.

\subsection{Effectiveness Evaluation}
Nevertheless, validating the calculated sustainability pillar, which could enhance the credibility of the trust score, is a complex task. This difficulty primarily stems from the absence of the ground truth, rendering quantitative analysis notably challenging. Therefore, this experiment analyzes and validates the effectiveness and value-adding properties of the sustainability pillar through a hypothetical case study.

\begin{table}[ht!]
\centering
\caption{The FL Configuration of the Proposals from the Two Branches}
\label{tab:case_config}
\resizebox{\columnwidth}{!}{%
\begin{tabular}{lll}
\toprule
\textit{Metric}                          & \textit{Proposal A}                               & \textit{Proposal B}     \\ \midrule
   \hline
Model                           & ConvNet2                                 & ConvNet2 \\   \hline
Local Rounds & 100                                      & 10  \\  \hline
Dataset                         & FEMNIST                                  & FEMNIST \\  \hline
\begin{tabular}[c]{@{}l@{}} Data Split \\ (Train, Val., Test)\end{tabular}     & 0.6/0.2/0.2                       &  0.6/0.2/0.2 \\  \hline
Batch Size                      & 50                                       & 50 \\  \hline
Loss                            & CrossEntropyLoss                         & CrossEntropyLoss \\  \hline
\begin{tabular}[c]{@{}l@{}} Consistent Label \\Distribution\end{tabular}   & False                                    & False \\  \hline
\rowcolor{gray!10} Number of Clients               & 1000                                 & 10  \\  \hline
\rowcolor{gray!10} Client Selection Rate           & 0.3                                     & 0.6\\  \hline
\rowcolor{gray!10} Federation Rounds       & 1000                                 & 10  \\  \hline
Clients Hardware                 & Intel i7-8650U                           & Intel i7-8650U   \\  \hline
Server Hardware                 & Intel i7-8650U                           & Intel i7-8650U   \\  \hline
\rowcolor{gray!10} Client Location                 & South Africa                             & Switzerland  \\  \hline
\rowcolor{gray!10} Server Location                 & South Africa                             & Switzerland  \\  \hline
Differential Privacy            & Epsilon 10                               & Epsilon 10 \\  \hline
Aggregation Method              & FedAvg                                   & FedAvg \\ \hline
\end{tabular}%
}
\end{table}

Assuming a multinational IT consulting company based in Luxembourg, with two research and development centers located in Zurich, Switzerland, and Johannesburg, South Africa. Both branches have simultaneously proposed an FL-based training proposal, with their respective training configurations outlined in the \tablename~\ref{tab:case_config}. However, due to limited resources, only one proposal can be implemented. As the director of the research and development centers, the decision-maker aims to follow the guidance of the AI-HLEG. It intends to evaluate the trust score of the two proposals using the algorithm proposed in this work. This calculation will ultimately determine which proposal should be adopted.

\tablename~\ref{tab:case_config} presents the configurations of the two proposals, which exhibit a high degree of similarity. The primary distinction lies in the fact that \textbf{Proposal A}, involving the Johannesburg team, necessitates a greater number of clients to participate in the training process and entails a substantially higher number of training rounds compared to \textbf{Proposal B}, which is proposed by the Zurich team. Both teams intend to conduct the training process at their local facilities. 

\begin{figure}[ht!]
\centering
\small
\subfloat{{\includegraphics[width=0.9\linewidth]{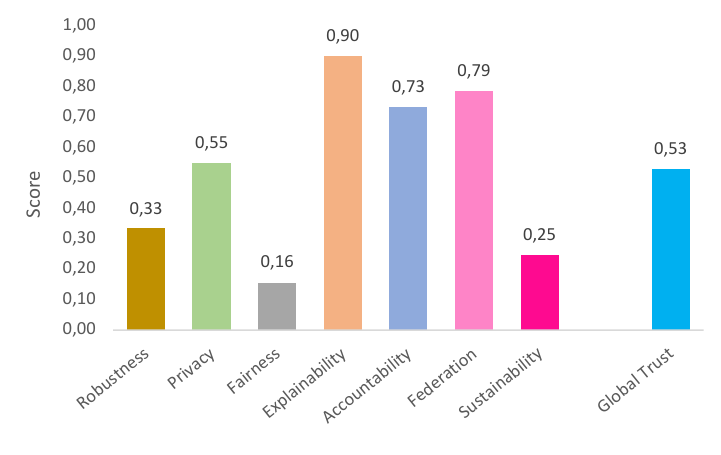}}}\\
\subfloat{{\includegraphics[width=0.9\linewidth]{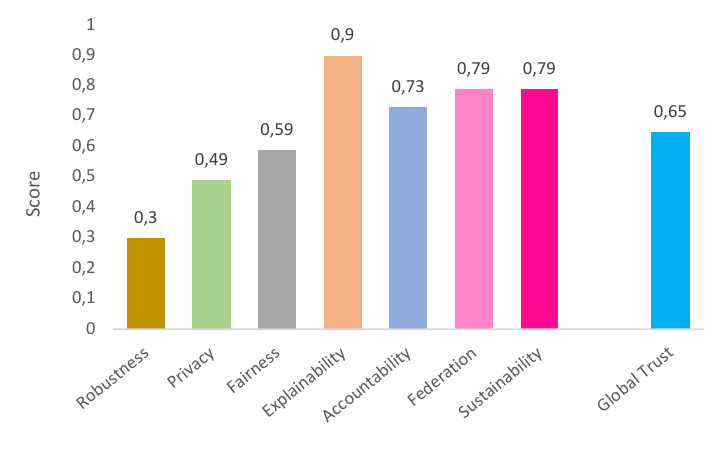}}}\\
  \caption{Results of Evaluation of the Proposed Algorithm for Proposal A (top) and Proposal B (bottom)}
    \label{fig:result_proposals}
\end{figure}

The director utilized the proposed system to upload the proposals submitted by the two teams. This system then computed and evaluated the scores of various pillars, such as robustness, privacy, and fairness, ultimately aggregating them to generate a trust score. This experiment assigned equal weight to all the pillars during calculations. 

The results of the system, as depicted in \figurename~\ref{fig:result_proposals}, indicate that both proposals have similar scores in different aspects, including explainability, accountability, and federation. This similarity can be attributed to the proximity of their respective configurations. As indicated in the \tablename~\ref{tab:notion_score_proposals}, both proposals demonstrated low levels of robustness as they were not optimized for resisting attacks. Regarding privacy, proposal B outperformed proposal A due to its significant number of nodes, which introduced more uncertainty and improved overall privacy. Besides, proposal B exhibited a greater fairness score compared to proposal A due to its superior level of client selection fairness, and the performance of the model among the clients is even.

\begin{table}[ht!]
\small
\centering
\caption{Sustainability Pillar and Notion Scores for Two Proposals}
\label{tab:notion_score_proposals}
{%
\begin{tabular}{lll}
\toprule
\textit{Metric}                      & \textit{Proposal A} & \textit{Proposal B} \\ \midrule
   \hline
\textbf{Sustainability Pillar}       & 0.25       & 0.79 \\ \hline
\rowcolor{gray!10}\begin{tabular}[c]{@{}l@{}}- Carbon Intensity of \\ Energy Grid Server\end{tabular}   & 0.11       & 0.98 \\ \hline
- Hardware Efficiency                  & 0.28       & 0.28 \\ \hline
\rowcolor{gray!10}- Federation Complexity                & 0.49       & 0.91 \\ \bottomrule    
\end{tabular}%
}
\end{table}

Before the inclusion of the sustainability pillar, the trust scores for the two proposals were relatively similar, with proposal A receiving a score of 0.58 and proposal B receiving a score of 0.63, indicating a minimal difference of 0.05. This posed a challenge in determining which proposal aligned more closely with the concept of trustworthiness. However, with the introduction of the sustainability pillar, the data presented in the \tablename~\ref{tab:notion_score_proposals} reveals that proposal B exhibited notable advantages regarding carbon intensity of energy source and federation complexity. As a result, the final trust scores were adjusted to 0.53 and 0.65 for proposal A and proposal B, respectively, resulting in an increased discrepancy of 0.12. Ultimately, proposal B emerged as the winner due to its superior performance in sustainability.

In summary, this experiment serves as a hypothetical case study to illustrate that the sustainability pillar effectively enhances users’ comprehension of the environmental impacts of FL systems and provides valuable support in decision-making. Moreover, it offers practical insights for the sustainable design and optimization of FL-based network management models, where energy efficiency, resource heterogeneity, and environmental constraints must be jointly considered.

\section{Discussion}
\label{sec:discussion}

The influence of individual metrics on CO$_2$eq emissions remains uncertain, yet all are currently weighted equally. For instance, training rounds and client count receive the same weight, despite differing environmental impacts. Hardware efficiency is currently estimated only from CPU/GPU benchmarks (e.g., PassMark), excluding RAM and embodied emissions from manufacturing, while indirect factors such as homomorphic encryption add further overhead.

Another limitation is the lack of reliable ground truth for emissions in FL. While direct measurement is difficult, recent testbeds with power consumption modules~\cite{feng2025practical} can approximate ground truth and enable real-world validation of the framework. In addition, this study primarily focused on model complexity, overlooking communication, which prior work~\cite{feng2025greendfl} shows contributes comparatively little to FL emissions. Future work will expand metrics to include communication overhead.

Experiments also show that electricity grid carbon intensity and device energy efficiency strongly affect sustainability. Thus, optimization algorithms that prioritize efficient devices in low-carbon regions while discouraging high-consumption nodes will be explored. Furthermore, the weights assigned to notions and metrics require refinement to better reflect their real-world importance.

Although computing sustainability metrics requires users to provide system-level factsheets, this typically involves metadata (e.g., hardware model, location) and poses limited privacy risks. Addressing these limitations will enhance the precision, coverage, and usability of sustainability assessments in FL systems.

\section{Summary and Future Work}
\label{sec:conclusion}

This work addresses mechanisms in the area of critical decision-making, particularly FL for network and service management as a privacy-preserving collaborative AI paradigm. It extends the taxonomy of trustworthy FL by introducing a sustainability pillar to assess the environmental impact of FL systems. Ten metrics are defined across three notions: hardware efficiency, federation complexity, and carbon intensity of the energy grid, and integrated into an evaluation algorithm that incorporates hardware specifications and geographic energy profiles of clients and servers. Experiments indicate that FL systems with lower model complexity, more efficient devices, and cleaner energy sources achieve higher trustworthiness scores. The proposed approach provides actionable insights for optimizing FL deployments under heterogeneous resources, distributed environments, and energy constraints.

Future work will refine the weighting of notions and metrics to better reflect actual carbon impact, extend the coverage of communication and network infrastructure overhead, and address trade-offs between pillars (e.g., privacy mechanisms increasing sustainability cost). In addition, validation on physical testbeds with power measurement modules will serve as a proxy for ground truth emissions. Finally, enhancing the prototype’s security, compatibility, and support for decentralized scenarios, as well as integrating metrics from the other six pillars, will further improve its comprehensiveness.

\section*{Acknowledgments}
This work has been partially supported by \textit{(a)} the Swiss Federal Office for Defense Procurement (armasuisse) with the CyberDFL project (CYD-C-2020003) and \textit{(b)} the University of Zürich (UZH).

\bibliographystyle{ieeetr}
\bibliography{acmart}

\end{document}